\documentclass[12pt,english]{article}
\usepackage{tgtermes}

\usepackage[T1]{fontenc}
\usepackage[latin9]{inputenc}
\usepackage[a4paper]{geometry}
\usepackage{rotfloat}
\usepackage{amsmath}
\usepackage{graphicx}
\usepackage[authoryear]{natbib}

\makeatletter

\providecommand{\tabularnewline}{\\}

\renewcommand{\textendash}{--}

\makeatother

\usepackage{babel}
\begin{document}

\title{On the Likelihood of Local Projection Models}

\author{Masahiro Tanaka\thanks{Department of Economics, Kanto Gakuen University; Graduate School
of Economics, Waseda University. Address: 200, Fujiagucho, Ohta, Gunma
169-8050 Japan. Email: gspddlnit45@toki.waseda.jp. Personal website:
https://sites.google.com/view/masahirotanakastat}}

\date{July 10, 2020}
\maketitle
\begin{abstract}
A local projection model is defined by a set of linear regressions
that account for the associations between exogenous variables and
an endogenous variable observed at different time points. While it
is standard practice to separately estimate individual regressions
using the ordinary least squares estimator, some recent studies treat
a local projection model as a multivariate regression with correlated
errors, i.e., seemingly unrelated regressions, and propose Bayesian
and non-Bayesian methods to improve the estimation accuracy. However,
it is not clear how and when this way of treatment of local projection
models is justified. The primary purpose of this paper is to fill
this gap by showing that the likelihood of local projection models
can be analytically derived from a stationary vector moving average
process. By means of numerical experiments, we confirm that this treatment
of local projections is tenable for finite samples.

\bigskip{}
\end{abstract}


\section{Introduction}

Local projections (LPs) \citep{Jorda2005} are linear regressions
that project observations of an endogenous variable at different time
periods, $y_{t},y_{t+1},...,y_{t+H}$, onto exogenous variables observed
at period $t$, $\boldsymbol{x}_{t}$:
\begin{equation}
y_{t+h}=\boldsymbol{\theta}_{\left(h\right)}^{\top}\boldsymbol{x}_{t-1}+u_{\left(h\right),t+h},\quad h=0,1,...,H;\;t=1,...,T,\label{eq: LP orig}
\end{equation}
where $\boldsymbol{\theta}_{\left(h\right)}$ is a coefficient vector
and $u_{\left(h\right),t+h}$ is a residual. $\boldsymbol{x}_{t-1}$
may include an intercept, preidentified structural shocks, the lags
of the endogenous variables up to period $t-1$, and other controls.
LPs are mainly employed for impulse response analysis. When the $n$th
element of $\boldsymbol{x}_{t-1}$, denoted by $x_{n,t-1}$, is a
preidentified structural shock, a sequence of the corresponding coefficients
is interpreted as an impulse response function, namely,
\[
\theta_{\left(h\right),n}=\partial y_{t+h}/\partial x_{n,t-1},\quad h=0,1,...,H.
\]
Because of their flexibility and robustness to model misspecification,
LPs are extensively applied to investigate the economic dynamics (e.g.,
\citealt{Ramey2016a} and cited therein). In this paper, we refer
to a set of LPs for the same endogenous variable (\ref{eq: LP orig})
as an LP model and statistical analysis using an LP model as the LP
method.

The LP method is intrinsically different from conventional approaches
to time series modeling. Standard time series models, such as the
vector autoregressive (VAR) model and its variants, are designed to
approximate the data generating process (DGP). In contrast, the LP
method does not explicitly assume the DGP and directly captures the
associations between the exogenous variables $\boldsymbol{x}_{t-1}$
and the endogenous variable observed at different time points $y_{t+h}$,
$h=0,1,...,H$.

It is standard practice to separately estimate LPs (\ref{eq: LP orig})
using the ordinary least squares (OLS) estimator, while \citet{Tanaka2020}
and \citet{El-Shagi2019} estimate an LP model as a multivariate regression
with correlated residuals, i.e., seemingly unrelated regressions (SUR)
\citep{Zellner1962}. \citet{Tanaka2020} and \citet{El-Shagi2019}
treat response variables for different projection horizons as correlated
but different variables and estimate an LP model as a multivariate
regression model where a vector of the responses $\boldsymbol{y}_{t}=\left(y_{t},y_{t+1},...,y_{t+H}\right)^{\top}$
are regressed on the covariates $\boldsymbol{x}_{t-1}$:

\begin{equation}
\boldsymbol{y}_{t}=\boldsymbol{\Theta}^{\top}\boldsymbol{x}_{t-1}+\boldsymbol{u}_{t},\quad t=1,...,T,\label{eq:LPs in SUR form}
\end{equation}
\[
\boldsymbol{u}_{t}=\left(u_{t},u_{t+1},...,u_{t+H}\right)^{\top},\quad t=1,...,T,
\]
\[
\boldsymbol{\Theta}=\left(\boldsymbol{\theta}_{\left(0\right)},\boldsymbol{\theta}_{\left(1\right)},...,\boldsymbol{\theta}_{\left(H\right)}\right).
\]
\citet{Tanaka2020} assumes that the residuals of an LP model are
distributed according to a multivariate normal distribution with the
covariance matrix $\boldsymbol{\Sigma}$, $\boldsymbol{u}_{t}\sim\mathcal{N}\left(\boldsymbol{0}_{H+1},\boldsymbol{\Sigma}\right)$,
where $\mathcal{N}\left(\boldsymbol{a},\boldsymbol{B}\right)$ denotes
a multivariate normal distribution with mean $\boldsymbol{a}$ and
covariance matrix $\boldsymbol{B}$. Thus, the (conditional) distribution
of $\boldsymbol{y}_{t}$ is specified as
\begin{equation}
\left(\boldsymbol{y}_{t}|\boldsymbol{x}_{t-1};\boldsymbol{\Theta},\boldsymbol{\Sigma}\right)\sim\mathcal{N}\left(\boldsymbol{\Theta}^{\top}\boldsymbol{x}_{t-1},\;\boldsymbol{\Sigma}\right),\quad t=1,...,T,\label{eq: dist LP orig}
\end{equation}
which leads to the following likelihood:
\begin{eqnarray}
p\left(\boldsymbol{Y}|\boldsymbol{\Theta},\boldsymbol{\Sigma},\boldsymbol{X}\right) & = & \prod_{t=1}^{T}p\left(\boldsymbol{y}_{t}|\boldsymbol{x}_{t-1};\boldsymbol{\Theta},\boldsymbol{\Sigma}\right)\nonumber \\
 & = & \prod_{t=1}^{T}f_{N}\left(\boldsymbol{y}_{t}|\boldsymbol{\Theta}^{\top}\boldsymbol{x}_{t-1},\boldsymbol{\Sigma}\right)\label{eq: SUR-like likelihood}\\
 & = & \left(2\pi\right)^{-\frac{\left(H+1\right)T}{2}}\left|\boldsymbol{\Sigma}\right|^{-\frac{T}{2}}\exp\left[-\frac{1}{2}\sum_{t=1}^{T}\left(\boldsymbol{y}_{t}-\boldsymbol{\Theta}^{\top}\boldsymbol{x}_{t-1}\right)^{\top}\boldsymbol{\Sigma}^{-1}\left(\boldsymbol{y}_{t}-\boldsymbol{\Theta}^{\top}\boldsymbol{x}_{t-1}\right)\right],\nonumber 
\end{eqnarray}
\[
\boldsymbol{Y}=\left(\boldsymbol{y}_{1},...,\boldsymbol{y}_{T}\right)^{\top},\quad\boldsymbol{X}=\left(\boldsymbol{x}_{1},...,\boldsymbol{x}_{T}\right)^{\top}.
\]
where $f_{N}\left(\boldsymbol{a}|\boldsymbol{b},\boldsymbol{C}\right)$
denotes the probability density function of a multivariate normal
distribution with mean $\boldsymbol{b}$ and covariance matrix $\boldsymbol{C}$
evaluated at $\boldsymbol{a}$. \citet{Tanaka2020} considers Bayesian
inference of LP models based on this likelihood. \citet{El-Shagi2019}
estimates LP models in a similar way to a feasible generalized least-squares
estimation of an SUR model, where the quadratic loss function that
he employs is closely related to the previously described likelihood
(\ref{eq: SUR-like likelihood}). \citet{Tanaka2020} and \citet{El-Shagi2019}
propose a Bayesian approach and a non-Bayesian approach to improve
the estimation of accuracy using a tailored prior and a penalty term,
respectively.

Whereas \citet{Tanaka2020} and \citet{El-Shagi2019} show that their
statistical approaches perform well by means of simulation studies,
they do not explicitly discuss how and when the likelihood of the
LP models that they utilize (\ref{eq: SUR-like likelihood}) is justified.
The primary focus of this paper is to fill this gap. We suggest that
their SUR-like treatment of LP models is statistically valid by showing
that the likelihood (\ref{eq: SUR-like likelihood}) can be analytically
derived from a stationary vector moving average (VMA) process with
the normality assumption.

\citet{Lusompa2019} proposes Bayesian and non-Bayesian methods to
estimate LPs. Unlike the SUR-like treatment in \citet{Tanaka2020}
and \citet{El-Shagi2019}, his approach requires estimation of all
the LPs for endogenous variables, i.e., the whole system of LPs; thus,
it is computationally demanding. Specifically, his Bayesian method
is not practical when the number of endogenous variables and/or the
number of projection horizons is large. In most situations, it is
not necessary to examine all the LPs because analysts' attention is
limited to coefficients that represent impulse responses of interest.
Thus, confirming the statistical validity of the SUR-like treatment
of LPs in \citet{Tanaka2020} and \citet{El-Shagi2019}, which is
more parsimonious than that of \citet{Lusompa2019}, has practical
importance.

We discuss the relationship between a VMA process and the SUR-like
likelihood of an LP model in two steps. First, we develop the entire
system of LPs from a stationary VMA process with normality assumption
(Section 2). Second, we define an LP model as a sub-system of the
entire system of LPs and derive its probabilistic representation,
i.e., the likelihood (Section 3). We see that the covariance matrix
of the residuals of an LP model can be represented as a (matrix-valued)
function of the parameters of the VMA process and that it is non-diagonal,
full rank, and symmetric positive definite. Therefore, the likelihood
of an LP model can be specified as in SUR. By means of numerical experiments,
we show that the derived representation of the residual covariance
matrix can be accurately estimated using the OLS estimator and Bayesian
Markov chain Monte Carlo method (Section 4). Therefore, the SUR-like
treatment of LP models is tenable for finite samples. 

\section{From VMA Process to Entire System of LPs }

This section derives an entire system of LPs that corresponds to the
underlying DGP specified by a VMA process. We assume that data $\left\{ \boldsymbol{w}_{t}\right\} $
are generated from an $L$th-order stationary VMA process:
\begin{equation}
\boldsymbol{w}_{t}=\boldsymbol{\varepsilon}_{t}+\sum_{l=1}^{L}\boldsymbol{\Gamma}_{l}\boldsymbol{\varepsilon}_{t-l},\quad\boldsymbol{\varepsilon}_{t}\sim\mathcal{N}\left(\boldsymbol{0}_{M},\boldsymbol{\Omega}_{\varepsilon}\right),\quad t=-\infty,...,T,\label{eq: VMA process}
\end{equation}
where $\boldsymbol{w}_{t}=\left(w_{t}^{\left\langle 1\right\rangle },...,w_{t}^{\left\langle M\right\rangle }\right)^{\top}$
is an $M$-dimensional vector of endogenous variables, $\boldsymbol{\varepsilon}_{t}$
is an $M$-dimensional vector of structural shocks that are distributed
according to a multivariate normal distribution with covariance matrix
$\boldsymbol{\Omega}_{\boldsymbol{\varepsilon}}$, and $\boldsymbol{\Gamma}_{l}$
is an $M$-by-$M$ VMA coefficient matrix. Without loss of generality,
intercepts are omitted. The process is not assumed to be fundamental;
thus, our discussion applies not only to VAR processes but also to
more general multivariate time series. The identification of structural
shocks is beyond the scope of this paper; impulse responses can be
estimated only when the shocks are exogenous or identified before
inference. 

An $h$ horizon LP is defined by a projection of $\boldsymbol{w}_{t+h}$
onto the lags of the endogenous variables, $\boldsymbol{w}_{t-1},...,\boldsymbol{w}_{t-L}$:
\begin{equation}
\boldsymbol{w}_{t+h}=\textrm{Proj}\left(\boldsymbol{w}_{t+h}|\boldsymbol{w}_{t-1},...,\boldsymbol{w}_{t-L}\right)+\boldsymbol{e}_{\left(h\right),t+h},\quad=0,1,...,H;\;t=1,...,T,\label{eq: whole system of LP}
\end{equation}
where $\textrm{Proj}\left(\boldsymbol{a}|\boldsymbol{b}\right)$ denotes
the orthogonal projection of $\boldsymbol{a}$ onto $\boldsymbol{b}$
and $\boldsymbol{e}_{\left(h\right),t+h}=\left(e_{\left(h\right),t+h}^{\left\langle 1\right\rangle },...,e_{\left(h\right),t+h}^{\left\langle M\right\rangle }\right)^{\top}$
is an $M$-dimensional vector of the LP residuals. The orthogonal
projection $\textrm{Proj}\left(\boldsymbol{w}_{t+h}|\boldsymbol{w}_{t-1},...,\boldsymbol{w}_{t-L}\right)$
is specified by the linear projection $\sum_{l=1}^{L}\left[\boldsymbol{\Phi}_{\left(h+1\right),l}\boldsymbol{w}_{t-l}\right]$,
where $\boldsymbol{\Phi}_{\left(h\right),l}$ is an $M$-by-$M$ LP
coefficient matrix whose $\left(m,n\right)$-element is denoted by
$\phi_{\left(h\right),l}^{\left\langle m\right\rangle \left\langle n\right\rangle }$.
Each LP is posed as

\[
\boldsymbol{w}_{t+h}=\sum_{l=1}^{L}\left[\boldsymbol{\Phi}_{\left(h+1\right),l}\boldsymbol{w}_{t-l}\right]+\boldsymbol{e}_{\left(h\right),t+h},\quad h=0,1,...,H;\;t=1,...,T.
\]
In the following section, we refer to this representation of LPs (or
equivalently (\ref{eq: whole system of LP})) as the entire system
of LPs, in distinction from an LP model for a specific endogenous
variable, such as (\ref{eq: LP orig}) and (\ref{eq:LPs in SUR form}).
Although multivariate time series data are considered, an LP model
is not a typical time series model. The purpose of the model is to
directly examine the relationship between variables measured at different
time points, not to recover the underlying DGP. 

The construction of the likelihood of LP models is different from
that of the standard multivariate time series models. For instance,
if the VMA process (\ref{eq: VMA process}) has a VAR representation,
the likelihood of the VAR model is written generically in the following
form:
\[
p\left(\boldsymbol{W}|\mathcal{P}_{VAR}\right)=\prod_{t}p\left(\boldsymbol{w}_{t}|\boldsymbol{w}_{t-1},...,\boldsymbol{w}_{t-L};\mathcal{P}_{VAR}\right),
\]
where $\mathcal{P}_{VAR}$ denotes a set of parameters that specify
the VAR model. Contrastingly, the likelihood of the LP representation
of the VMA process (\ref{eq: whole system of LP}) is specified as
\[
p\left(\boldsymbol{W}|\mathcal{P}_{LP}\right)=\prod_{m=1}^{M}\prod_{t=1}^{T}\prod_{h=0}^{H}p\left(w_{t+h}^{\left\langle m\right\rangle }|\boldsymbol{w}_{t-1},...,\boldsymbol{w}_{t-L};\mathcal{P}_{LP}\right),
\]
where $\mathcal{P}_{LP}$ denotes a set of parameters that specify
the entire system of LPs. As shown in \citet{Jorda2005}, when the
true DGP is a VAR process, the LP coefficients $\boldsymbol{\Phi}_{\left(h+1\right),l}$,
$l=1,...,L$, are parametrized by the coefficients that specify the
VAR process. 

The residuals of the entire system of LPs $\boldsymbol{e}_{\left(h\right),t+h}$
have autocorrelations. \citet{Lusompa2019} shows that the process
of $\boldsymbol{e}_{\left(h\right),t+h}$ is known: for $t=1,...,T$,
\[
\boldsymbol{e}_{\left(h\right),t+h}=\begin{cases}
\boldsymbol{v}_{\left(h\right),t+h}, & h=0,\\
\sum_{i=1}^{h}\left[\boldsymbol{\Gamma}_{i}\boldsymbol{\varepsilon}_{t+h-i}\right]+\boldsymbol{v}_{\left(h\right),t+h}, & h=1,...,H,
\end{cases}
\]
where $\boldsymbol{v}_{\left(h\right),t+h}$ is an $M$-dimensional
vector of serially uncorrelated residuals. In the population, it follows
that
\begin{eqnarray*}
\boldsymbol{v}_{\left(h\right),t+h} & = & \boldsymbol{\varepsilon}_{t+h},\quad h=0,1,...,H;\;t=1,...,T,\\
\boldsymbol{\Phi}_{\left(h\right),1} & = & \boldsymbol{\Gamma}_{h},\quad h=1,...,H.
\end{eqnarray*}
The residuals are written as
\begin{eqnarray}
\boldsymbol{e}_{\left(h\right),t+h} & = & \begin{cases}
\boldsymbol{\varepsilon}_{t}, & h=0,\\
\sum_{i=1}^{h}\left[\boldsymbol{\Phi}_{\left(h\right),1}\boldsymbol{\varepsilon}_{t+h-i}\right]+\boldsymbol{\varepsilon}_{t+h}, & h=1,...,H,
\end{cases}\label{eq: process of LP residuals}
\end{eqnarray}
for $t=1,...,T$.

\section{From Entire System of LPs to LP model}

An LP model for the $m$th endogenous variable is defined by extracting
a portion of the LPs, or a sub-system, from the remainder of the entire
system:
\[
w_{t+h}^{\left\langle m\right\rangle }=\sum_{l=1}^{L}\left[\left(\boldsymbol{\Phi}_{l}^{\left\langle m\right\rangle }\right)^{\top}\boldsymbol{w}_{t-l}\right]+e_{\left(h\right),t+h}^{\left\langle m\right\rangle },\quad h=0,1,...,H;\;t=1,...,T,
\]
\[
\boldsymbol{\Phi}_{l}^{\left\langle m\right\rangle }=\left(\boldsymbol{\phi}_{\left(0\right),l}^{\left\langle m\right\rangle },\boldsymbol{\phi}_{\left(1\right),l}^{\left\langle m\right\rangle },...,\boldsymbol{\phi}_{\left(H\right),l}^{\left\langle m\right\rangle }\right),\quad l=1,...,L,
\]
where $\boldsymbol{\phi}_{\left(h\right),l}^{\left\langle m\right\rangle }$
is a vector containing the element of the $m$th row of $\boldsymbol{\Phi}_{\left(h\right),l}$.
This sub-system is summarized in an analogous fashion to (\ref{eq:LPs in SUR form}):
for $t=1,...,T$,

\[
\boldsymbol{w}_{t}^{\left\langle m\right\rangle }=\left(\boldsymbol{\Phi}^{\left\langle m\right\rangle }\right)^{\top}\tilde{\boldsymbol{w}}_{t-1}+\boldsymbol{e}_{t}^{\left\langle m\right\rangle },
\]
\[
\boldsymbol{w}_{t}^{\left\langle m\right\rangle }=\left(w_{t}^{\left\langle m\right\rangle },w_{t+1}^{\left\langle m\right\rangle },...,w_{t+H}^{\left\langle m\right\rangle }\right)^{\top},\quad t=1,...,T,
\]

\[
\tilde{\boldsymbol{w}}_{t-1}=\left(\boldsymbol{w}_{t-1}^{\top},...,\boldsymbol{w}_{t-L}^{\top}\right)^{\top},
\]

\[
\boldsymbol{e}_{t}^{\left\langle m\right\rangle }=\left(e_{t}^{\left\langle m\right\rangle },e_{t+1}^{\left\langle m\right\rangle },...,e_{t+H}^{\left\langle m\right\rangle }\right)^{\top},
\]
\[
\boldsymbol{\Phi}^{\left\langle m\right\rangle }=\left(\begin{array}{c}
\boldsymbol{\Phi}_{1}^{\left\langle m\right\rangle }\\
\vdots\\
\boldsymbol{\Phi}_{L}^{\left\langle m\right\rangle }
\end{array}\right),
\]
The likelihood of the sub-system is structured as
\begin{eqnarray}
p\left(\boldsymbol{W}^{\left\langle m\right\rangle }|\mathcal{P}_{LP}\right) & = & \prod_{t=1}^{T}\prod_{h=0}^{H}p\left(w_{t+h}^{\left\langle m\right\rangle }|\boldsymbol{w}_{t-1},...,\boldsymbol{w}_{t-L};\mathcal{P}_{LP}\right)\label{eq: LP likelihood}\\
 & = & \prod_{t=1}^{T}p\left(\boldsymbol{w}_{t}^{\left\langle m\right\rangle }|\boldsymbol{w}_{t-1},...,\boldsymbol{w}_{t-L};\mathcal{P}_{LP}\right),\nonumber 
\end{eqnarray}
\[
\boldsymbol{W}^{\left\langle m\right\rangle }=\left(\boldsymbol{w}_{1}^{\left\langle m\right\rangle },...,\boldsymbol{w}_{T}^{\left\langle m\right\rangle }\right)^{\top}.
\]
This expression corresponds to (\ref{eq: SUR-like likelihood}).

In the following section, we show that the residuals of the sub-system
are distributed according to a multivariate distribution, $\boldsymbol{e}_{t}^{\left\langle m\right\rangle }\sim\mathcal{N}\left(\boldsymbol{0}_{H+1},\boldsymbol{\Xi}\right)$,
where $\boldsymbol{\Xi}$ is a symmetric positive definite matrix,
and that the distribution of $\boldsymbol{w}_{t}^{\left\langle m\right\rangle }$
is specified as

\begin{equation}
\left(\boldsymbol{w}_{t}^{\left\langle m\right\rangle }|\tilde{\boldsymbol{w}}_{t-1};\boldsymbol{\Phi}^{\left\langle m\right\rangle },\boldsymbol{\Sigma}\right)\sim\mathcal{N}\left(\left(\boldsymbol{\Phi}^{\left\langle m\right\rangle }\right)^{\top}\tilde{\boldsymbol{w}}_{t-1},\;\boldsymbol{\Xi}\right),\quad t=1,...,T.\label{eq: dist LP alt}
\end{equation}
This expression is a mirror image of (\ref{eq: dist LP orig}). We
analytically derive the covariance of the residuals of the sub-system
$\boldsymbol{e}_{t}^{\left\langle m\right\rangle }$ as a matrix-valued
function of the parameters of the LP coefficients for the $m$th endogenous
variable and the covariance of the structural shocks $\boldsymbol{\Omega}_{\boldsymbol{\varepsilon}}$,
which shows that $\boldsymbol{\Xi}$ is proper as a covariance matrix
because it is full rank, symmetric, and positive definite. For readers'
convenience, the correspondence between the notations in the SUR-like
treatment of an LP model (\ref{eq: SUR-like likelihood}) and the
VMA process is summarized in Table 1.

From (\ref{eq: process of LP residuals}), using the LP coefficients
and the structural shocks, the residuals can be rewritten as follows:

\begin{eqnarray*}
\boldsymbol{e}_{t}^{\left\langle m\right\rangle } & = & \left(\begin{array}{c}
e_{\left(0\right),t}^{\left\langle m\right\rangle }\\
e_{\left(1\right),t+1}^{\left\langle m\right\rangle }\\
e_{\left(2\right),t+2}^{\left\langle m\right\rangle }\\
\vdots\\
e_{\left(H\right),t+H}^{\left\langle m\right\rangle }
\end{array}\right)\\
 & = & \left(\begin{array}{c}
\varepsilon_{t}^{\left\langle m\right\rangle }\\
\left(\boldsymbol{\phi}_{\left(1\right),1}^{\left\langle m\right\rangle }\right)^{\top}\boldsymbol{\varepsilon}_{t}+\varepsilon_{t+1}^{\left\langle m\right\rangle }\\
\left(\boldsymbol{\phi}_{\left(2\right),1}^{\left\langle m\right\rangle }\right)^{\top}\boldsymbol{\varepsilon}_{t}+\left(\boldsymbol{\phi}_{\left(1\right),1}^{\left\langle m\right\rangle }\right)^{\top}\boldsymbol{\varepsilon}_{t+1}+\varepsilon_{t+2}^{\left\langle m\right\rangle }\\
\vdots\\
\left(\boldsymbol{\phi}_{\left(H\right),1}^{\left\langle m\right\rangle }\right)^{\top}\boldsymbol{\varepsilon}_{t}+\left(\boldsymbol{\phi}_{\left(H-1\right),1}^{\left\langle m\right\rangle }\right)^{\top}\boldsymbol{\varepsilon}_{t+1}+\cdots+\left(\boldsymbol{\phi}_{\left(1\right),1}^{\left\langle m\right\rangle }\right)^{\top}\boldsymbol{\varepsilon}_{t+H-1}+\varepsilon_{t+H}^{\left\langle m\right\rangle }
\end{array}\right).
\end{eqnarray*}
Let $\boldsymbol{\iota}_{m}$ denote an $M$-dimensional vector of
zeros, where the $m$th element is replaced by one. The residuals
are arranged as
\begin{eqnarray*}
\boldsymbol{e}_{t}^{\left\langle m\right\rangle } & = & \left(\begin{array}{c}
\boldsymbol{\iota}_{m}^{\top}\boldsymbol{\varepsilon}_{t}\\
\left(\boldsymbol{\phi}_{\left(1\right),1}^{\left\langle m\right\rangle }\right)^{\top}\boldsymbol{\varepsilon}_{t}+\boldsymbol{\iota}_{m}^{\top}\boldsymbol{\varepsilon}_{t+1}\\
\left(\boldsymbol{\phi}_{\left(2\right),1}^{\left\langle m\right\rangle }\right)^{\top}\boldsymbol{\varepsilon}_{t}+\left(\boldsymbol{\phi}_{\left(1\right),1}^{\left\langle m\right\rangle }\right)^{\top}\boldsymbol{\varepsilon}_{t+1}+\boldsymbol{\iota}_{m}^{\top}\boldsymbol{\varepsilon}_{t+2}\\
\vdots\\
\left(\boldsymbol{\phi}_{\left(H\right),1}^{\left\langle m\right\rangle }\right)^{\top}\boldsymbol{\varepsilon}_{t}+\left(\boldsymbol{\phi}_{\left(H-1\right),1}^{\left\langle m\right\rangle }\right)^{\top}\boldsymbol{\varepsilon}_{t+1}+\cdots+\left(\boldsymbol{\phi}_{\left(1\right),1}^{\left\langle m\right\rangle }\right)^{\top}\boldsymbol{\varepsilon}_{t+H-1}+\boldsymbol{\iota}_{m}^{\top}\boldsymbol{\varepsilon}_{t+H}
\end{array}\right)\\
 & = & \left(\begin{array}{ccccc}
\boldsymbol{\iota}_{m}^{\top}\\
\left(\boldsymbol{\phi}_{\left(1\right),1}^{\left\langle m\right\rangle }\right)^{\top} & \boldsymbol{\iota}_{m}^{\top}\\
\left(\boldsymbol{\phi}_{\left(2\right),1}^{\left\langle m\right\rangle }\right)^{\top} & \left(\boldsymbol{\phi}_{\left(1\right),1}^{\left\langle m\right\rangle }\right)^{\top} & \boldsymbol{\iota}_{m}^{\top}\\
\vdots & \vdots & \cdot & \ddots\\
\left(\boldsymbol{\phi}_{\left(H\right),1}^{\left\langle m\right\rangle }\right)^{\top} & \left(\boldsymbol{\phi}_{\left(H-1\right),1}^{\left\langle m\right\rangle }\right)^{\top} & \cdots & \left(\boldsymbol{\phi}_{\left(1\right),1}^{\left\langle m\right\rangle }\right)^{\top} & \boldsymbol{\iota}_{m}^{\top}
\end{array}\right)\left(\begin{array}{c}
\boldsymbol{\varepsilon}_{t}\\
\boldsymbol{\varepsilon}_{t+1}\\
\boldsymbol{\varepsilon}_{t+2}\\
\vdots\\
\boldsymbol{\varepsilon}_{t+H}
\end{array}\right).
\end{eqnarray*}
The covariance of $\boldsymbol{e}_{t}^{\left\langle m\right\rangle }$
is represented using the LP coefficients and the covariance matrix
of the structural shocks as
\begin{equation}
\boldsymbol{\Xi}=\bar{\boldsymbol{\Phi}}^{\left\langle m\right\rangle }\left(\boldsymbol{I}_{H+1}\otimes\boldsymbol{\Omega}_{\varepsilon}\right)\left(\bar{\boldsymbol{\Phi}}^{\left\langle m\right\rangle }\right)^{\top},\label{eq: matSig_LP}
\end{equation}
\begin{equation}
\bar{\boldsymbol{\Phi}}^{\left\langle m\right\rangle }=\left(\begin{array}{ccccc}
\boldsymbol{\iota}_{m}^{\top}\\
\left(\boldsymbol{\phi}_{\left(1\right),1}^{\left\langle m\right\rangle }\right)^{\top} & \boldsymbol{\iota}_{m}^{\top}\\
\left(\boldsymbol{\phi}_{\left(2\right),1}^{\left\langle m\right\rangle }\right)^{\top} & \left(\boldsymbol{\phi}_{\left(1\right),1}^{\left\langle m\right\rangle }\right)^{\top} & \boldsymbol{\iota}_{m}^{\top}\\
\vdots & \vdots & \cdot & \ddots\\
\left(\boldsymbol{\phi}_{\left(H\right),1}^{\left\langle m\right\rangle }\right)^{\top} & \left(\boldsymbol{\phi}_{\left(H-1\right),1}^{\left\langle m\right\rangle }\right)^{\top} & \cdots & \left(\boldsymbol{\phi}_{\left(1\right),1}^{\left\langle m\right\rangle }\right)^{\top} & \boldsymbol{\iota}_{m}^{\top}
\end{array}\right).\label{eq: matPhi_bar}
\end{equation}
This expression implies that $\boldsymbol{\Xi}$ is non-diagonal,
full rank, and symmetric positive definite. Thus, $\boldsymbol{\Xi}$
is properly defined as a covariance matrix, and the distribution of
$\boldsymbol{w}_{t}^{\left\langle m\right\rangle }$ can be written
in the form of (\ref{eq: dist LP alt}). 

Given $\boldsymbol{\Omega}_{\boldsymbol{\varepsilon}}$, the likelihood
of this sub-system is independent from the VMA coefficients for the
remainder of the whole system (\ref{eq: whole system of LP}), because
$\bar{\boldsymbol{\Phi}}^{\left\langle m\right\rangle }$ does not
contain the LP coefficients for the remaining endogenous variables,
namely, $\boldsymbol{\phi}_{\left(h\right),1}^{\left\langle m^{\prime}\right\rangle }$,
$h=0,1,...,H$; $m^{\prime}\neq m$. Therefore, to recover the LP
coefficients from the data, it is sufficient to estimate the sub-system.
The likelihood (\ref{eq: LP likelihood}) is specified as follows:

\begin{eqnarray*}
p\left(\boldsymbol{W}^{\left\langle m\right\rangle }|\boldsymbol{\Phi}^{\left\langle m\right\rangle },\boldsymbol{\Xi}\right) & = & \prod_{t=1}^{T}p\left(\boldsymbol{w}_{t}^{\left\langle m\right\rangle }|\tilde{\boldsymbol{w}}_{t-1};\boldsymbol{\Phi}^{\left\langle m\right\rangle },\boldsymbol{\Xi}\right)\\
 & = & \prod_{t=1}^{T}f_{N}\left(\boldsymbol{w}_{t}^{\left\langle m\right\rangle }|\left(\boldsymbol{\Phi}^{\left\langle m\right\rangle }\right)^{\top}\tilde{\boldsymbol{w}}_{t-1},\;\boldsymbol{\Xi}\right)\\
 & = & \left(2\pi\right)^{-\frac{\left(H+1\right)T}{2}}\left|\boldsymbol{\Xi}\right|^{-\frac{T}{2}}\exp\left[-\frac{1}{2}\sum_{t=1}^{T}\left(\boldsymbol{w}_{t}^{\left\langle m\right\rangle }-\left(\boldsymbol{\Phi}^{\left\langle m\right\rangle }\right)^{\top}\tilde{\boldsymbol{w}}_{t-1}\right)^{\top}\right.\\
 &  & \hspace{150bp}\left.\boldsymbol{\Xi}^{-1}\left(\boldsymbol{w}_{t}^{\left\langle m\right\rangle }-\left(\boldsymbol{\Phi}^{\left\langle m\right\rangle }\right)^{\top}\tilde{\boldsymbol{w}}_{t-1}\right)\right]
\end{eqnarray*}
There is a correspondence between this likelihood presentation and
the SUR-like likelihood (\ref{eq: SUR-like likelihood}). 

From the foregoing analysis, we confirm that the probabilistic representations
of an LP model in (\ref{eq: dist LP alt}), and thus, (\ref{eq: dist LP orig})
are justifiable when the true DGP is a stationary VMA process with
the normality assumption and that the likelihood of an LP model based
on the SUR-like treatment (\ref{eq: SUR-like likelihood}), which
is applied in \citet{Tanaka2020}, is statistically valid.

\citet{Lusompa2019} proposes a Bayesian method to estimate the entire
system of LPs (\ref{eq: whole system of LP}). His posterior simulation
algorithm consists of two steps. First, the horizon 0 LP, which is
equivalent to a VAR model, is estimated. Second, for each draw, the
remaining coefficients are simulated from the conditional posteriors
sequentially for $h=1,2,...,H$. This algorithm is computationally
demanding, because all the coefficient matrices of the entire system
are inferred. In most situations, the analyst's focus is limited to
a small portion of these matrices, e.g., coefficients that represent
impulse responses of interest. Thus, inference about the whole system
is unnecessary. As long as structural shocks of interest are identified
before inference, the SUR-like treatment of LPs is advantageous to
the approach of \citet{Lusompa2019}. 

\section{Numerical Illustration}

By numerical experiments, we investigate the small-sample performance
of non-Bayesian and Bayesian approaches in the inference of an LP
model. Synthetic data are generated from the following VMA process:

\[
\boldsymbol{w}_{t}=\boldsymbol{\varepsilon}_{t}+\sum_{l=1}^{L}\boldsymbol{\Gamma}_{l}\boldsymbol{\varepsilon}_{t-l},\quad\boldsymbol{\varepsilon}_{t}\sim\mathcal{N}\left(\boldsymbol{0}_{M},\boldsymbol{\Omega}_{\varepsilon}\right).
\]
We fix $M=3$, $L=5$, $H=7$, and $T=200$. Our focus is inference
of the responses of the second element of $\boldsymbol{w}_{t}$ to
the first element of $\boldsymbol{w}_{t}$, which is exogenous to
the system. A covariate vector $\boldsymbol{x}_{t}$ includes $L$
lags of $\boldsymbol{w}_{t}$ and an intercept. The parameters for
the synthetic data are randomly generated. Let $\gamma_{i,j,l}$ denote
the $\left(i,j\right)$-element of $\boldsymbol{\Gamma}_{l}.$ The
first element of $\boldsymbol{w}_{t}$ is a preidentified structural
shock $w_{1,t}=\varepsilon_{1,t}$; thus, all the first rows of $\boldsymbol{\Gamma}_{l}$
are zero. A sequence of the $\left(2,1\right)$-elements of $\boldsymbol{\Gamma}_{l}$,
which is denoted by $\boldsymbol{\gamma}_{IRF}=\left\{ \gamma_{2,1,1},...,\gamma_{2,1,L}\right\} $,
represents the impulse responses of interest. The elements of $\boldsymbol{\gamma}_{IRF}$
are specified as
\[
\gamma_{2,1,l}=\frac{l\exp\left(d\left(1-l\right)\right)}{\sum_{l=1}^{L}l\exp\left(d\left(1-l\right)\right)},\quad l=1,...,L,
\]
where $d$ is a parameter that controls the shape of the impulse response
function. We fix $d=0.8$. The diagonal elements of $\boldsymbol{\Gamma}_{1}$
are uniformly sampled from the unit interval, $\left(0,1\right)$.
The second to third rows of $\boldsymbol{\Gamma}_{l}$ are set to
$\boldsymbol{\Gamma}_{2:3,1:3,l}=0.5\left(L-l\right)L^{-1}\boldsymbol{\Gamma}^{*}$,
where each entry of $\boldsymbol{\Gamma}^{*}$ is independently generated
from the standard normal distribution. The covariance matrix of $\boldsymbol{\varepsilon}_{t}$
is specified by $\boldsymbol{\Omega}_{\varepsilon}=\textrm{blkdiag}\left(\omega_{1}^{2},\boldsymbol{\Omega}_{\boldsymbol{\varepsilon}}^{*}\right)$.
$\omega_{1}^{2}$ is simulated from an inverse gamma distribution
with a shape parameter of 1 and scale parameter of 1, while $\boldsymbol{\Omega}_{\boldsymbol{\varepsilon}}^{*}$
is simulated from an inverse Wishart distribution with scale matrix
$\boldsymbol{I}_{2}$ and $2$ degrees of freedom. 

First, we consider a feasible generalized least-squares estimator,
as discussed in \citet{El-Shagi2019}. In the first step, $\boldsymbol{\Theta}$
is estimated using the OLS estimator,
\[
\hat{\boldsymbol{\Theta}}=\left(\boldsymbol{X}^{\top}\boldsymbol{X}\right)^{-1}\boldsymbol{X}^{\top}\boldsymbol{Y},
\]
where $\boldsymbol{Y}=\left(\boldsymbol{y}_{1},...,\boldsymbol{y}_{T}\right)^{\top}$
and $\boldsymbol{X}=\left(\boldsymbol{x}_{0},\boldsymbol{x}_{1},...,\boldsymbol{x}_{T-1}\right)^{\top}$.
Second, $\boldsymbol{\Sigma}$ is estimated based on the OLS residuals,
\[
\hat{\boldsymbol{\Sigma}}=T^{-1}\left(\boldsymbol{Y}-\boldsymbol{X}\hat{\boldsymbol{\Theta}}\right)^{\top}\left(\boldsymbol{Y}-\boldsymbol{X}\hat{\boldsymbol{\Theta}}\right).
\]
The second-step estimate of $\boldsymbol{\Theta}$ is obtained using
$\hat{\boldsymbol{\Sigma}}$. The system might be re-estimated iteratively.
The validity of this approach hinges critically on the precision of
$\hat{\boldsymbol{\Sigma}}$. The parameters $\boldsymbol{\Gamma}_{l}$
and $\boldsymbol{\Sigma}_{\boldsymbol{\varepsilon}}$ are generated
once and are fixed throughout the experiment, while the structural
shocks $\boldsymbol{\varepsilon}_{t}$ are randomly generated for
each trial. Figures 1\textendash 3 compare $\hat{\boldsymbol{\Sigma}}$
for 1,000 synthetic data with the true value of $\boldsymbol{\Sigma}$
given by (\ref{eq: matSig_LP}) and (\ref{eq: matPhi_bar}). Vertical
lines denote the true values, and the histograms represent the OLS
estimates of $\boldsymbol{\Sigma}$ for different data. As evident
from these figures, $\boldsymbol{\Sigma}$ can be consistently estimated
based on the OLS residuals. Figure 4 shows histograms of the estimates
of $\boldsymbol{\gamma}_{IRF}$. We see that $\boldsymbol{\gamma}_{IRF}$
can be estimated by the OLS estimator. 

Next, in a similar vein, we consider the Bayesian inference of an
LP model. To minimize the influence of priors on the posterior, we
employ a flat prior for $\boldsymbol{\Theta}$, $p\left(\boldsymbol{\Theta}\right)\propto1$,
and a scale-invariant Jeffreys prior for $\boldsymbol{\Sigma}$, $p\left(\boldsymbol{\Sigma}\right)\propto\left|\boldsymbol{\Sigma}\right|^{-\left(H+2\right)/2}$.
Posterior draws are simulated using a Gibbs sampler, and the conditionals
are specified as
\[
\textrm{vec}\left(\boldsymbol{\Theta}\right)|\boldsymbol{\Sigma},\boldsymbol{Y},\boldsymbol{X}\sim\mathcal{N}\left(\boldsymbol{m},\boldsymbol{P}^{-1}\right),
\]
\[
\boldsymbol{m}=\boldsymbol{P}^{-1}\left(\boldsymbol{\Sigma}^{-1}\otimes\boldsymbol{X}^{\top}\right)\textrm{vec}\left(\boldsymbol{Y}\right),
\]
\[
\boldsymbol{P}=\boldsymbol{\Sigma}^{-1}\otimes\boldsymbol{X}^{\top}\boldsymbol{X},
\]
\[
\boldsymbol{\Sigma}|\boldsymbol{\Theta},\boldsymbol{Y},\boldsymbol{X}\sim\mathcal{IW}\left(T,\;\left(\boldsymbol{Y}-\boldsymbol{X}\boldsymbol{\Theta}\right)^{\top}\left(\boldsymbol{Y}-\boldsymbol{X}\boldsymbol{\Theta}\right)\right),
\]
where $\mathcal{IW}\left(a,\boldsymbol{B}\right)$ denotes an inverse
Wishart distribution with $a$ degrees of freedom and scale matrix
$\boldsymbol{B}$. We generate 11,000 draws and use the last 10,000
draws for the posterior analysis. All the chains in the experiments
passed the \citet{Geweke1992} test at the 5 percent level. Figures
5\textendash 7 contain histograms of the posterior mean estimates
of $\boldsymbol{\Sigma}$. As with the OLS estimator, the posterior
means of $\boldsymbol{\Sigma}$ are distributed around the true values.
Figure 8 depicts histograms of the posterior mean estimates of the
impulse responses of interest. The estimated impulse responses are
distributed around the true values. Therefore, Bayesian analysis based
on the SUR-like treatment of LP models is tenable for finite samples. 

\bibliographystyle{econ}
\bibliography{reference}

\clearpage{}

\begin{sidewaystable}
\caption{Correspondence of notations}

\medskip{}

\centering{}%
\begin{tabular}{lccc}
\hline 
Description & Notations in LP model  &  & Notations in VMA process\tabularnewline
\hline 
Endogenous variable of interest & $y_{t+h}$ & $\Leftrightarrow$ & $w_{t+h}^{\left\langle m\right\rangle }$\tabularnewline
 &  &  & \tabularnewline
Vector of endogenous variable of interest & $\boldsymbol{y}_{t}=\left(y_{t},y_{t+1},...,y_{t+H}\right)^{\top}$ & $\Leftrightarrow$ & $\boldsymbol{w}_{t}^{\left\langle m\right\rangle }=\left(w_{t}^{\left\langle m\right\rangle },w_{t+1}^{\left\langle m\right\rangle },...,w_{t+H}^{\left\langle m\right\rangle }\right)^{\top}$\tabularnewline
 &  &  & \tabularnewline
Vector of exogenous variables & $\boldsymbol{x}_{t-1}$ & $\Leftrightarrow$ & $\tilde{\boldsymbol{w}}_{t-1}=\left(\boldsymbol{w}_{t-1}^{\top},...,\boldsymbol{w}_{t-L}^{\top}\right)^{\top}$\tabularnewline
 &  &  & \tabularnewline
Vector of LP residuals & $\boldsymbol{u}_{t}=\left(u_{\left(0\right),t+0},u_{\left(1\right),t+1},...,u_{\left(H\right),t+H}\right)^{\top}$ & $\Leftrightarrow$ & $\boldsymbol{e}_{t}^{\left\langle m\right\rangle }=\left(e_{\left(0\right),t}^{\left\langle m\right\rangle },e_{\left(1\right),t+1}^{\left\langle m\right\rangle },...,e_{\left(H\right),t+H}^{\left\langle m\right\rangle }\right)^{\top}$\tabularnewline
 &  &  & \tabularnewline
Matrix of LP coefficients & $\boldsymbol{\Theta}=\left(\boldsymbol{\theta}_{\left(0\right)},\boldsymbol{\theta}_{\left(1\right)},...,\boldsymbol{\theta}_{\left(H\right)}\right)$ & $\Leftrightarrow$ & $\boldsymbol{\Phi}^{\left\langle m\right\rangle }=\left(\begin{array}{c}
\boldsymbol{\Phi}_{1}^{\left\langle m\right\rangle }\\
\vdots\\
\boldsymbol{\Phi}_{L}^{\left\langle m\right\rangle }
\end{array}\right)=\left(\begin{array}{cccc}
\boldsymbol{\phi}_{\left(0\right),1}^{\left\langle m\right\rangle } & \boldsymbol{\phi}_{\left(1\right),1}^{\left\langle m\right\rangle } & \cdots & \boldsymbol{\phi}_{\left(H\right),L}^{\left\langle m\right\rangle }\\
\vdots & \vdots & \cdot & \vdots\\
\boldsymbol{\phi}_{\left(0\right),L}^{\left\langle m\right\rangle } & \boldsymbol{\phi}_{\left(1\right),L}^{\left\langle m\right\rangle } & \cdots & \boldsymbol{\phi}_{\left(H\right),L}^{\left\langle m\right\rangle }
\end{array}\right)$\tabularnewline
 &  &  & \tabularnewline
Covariance matrix of LP residuals & $\boldsymbol{\Sigma}$ & $\Leftrightarrow$ & $\boldsymbol{\Xi}$\tabularnewline
\hline 
\end{tabular}
\end{sidewaystable}

\clearpage{}

\begin{figure}
\caption{OLS estimates of $\boldsymbol{\Sigma}$ (1)}

\begin{centering}
\includegraphics{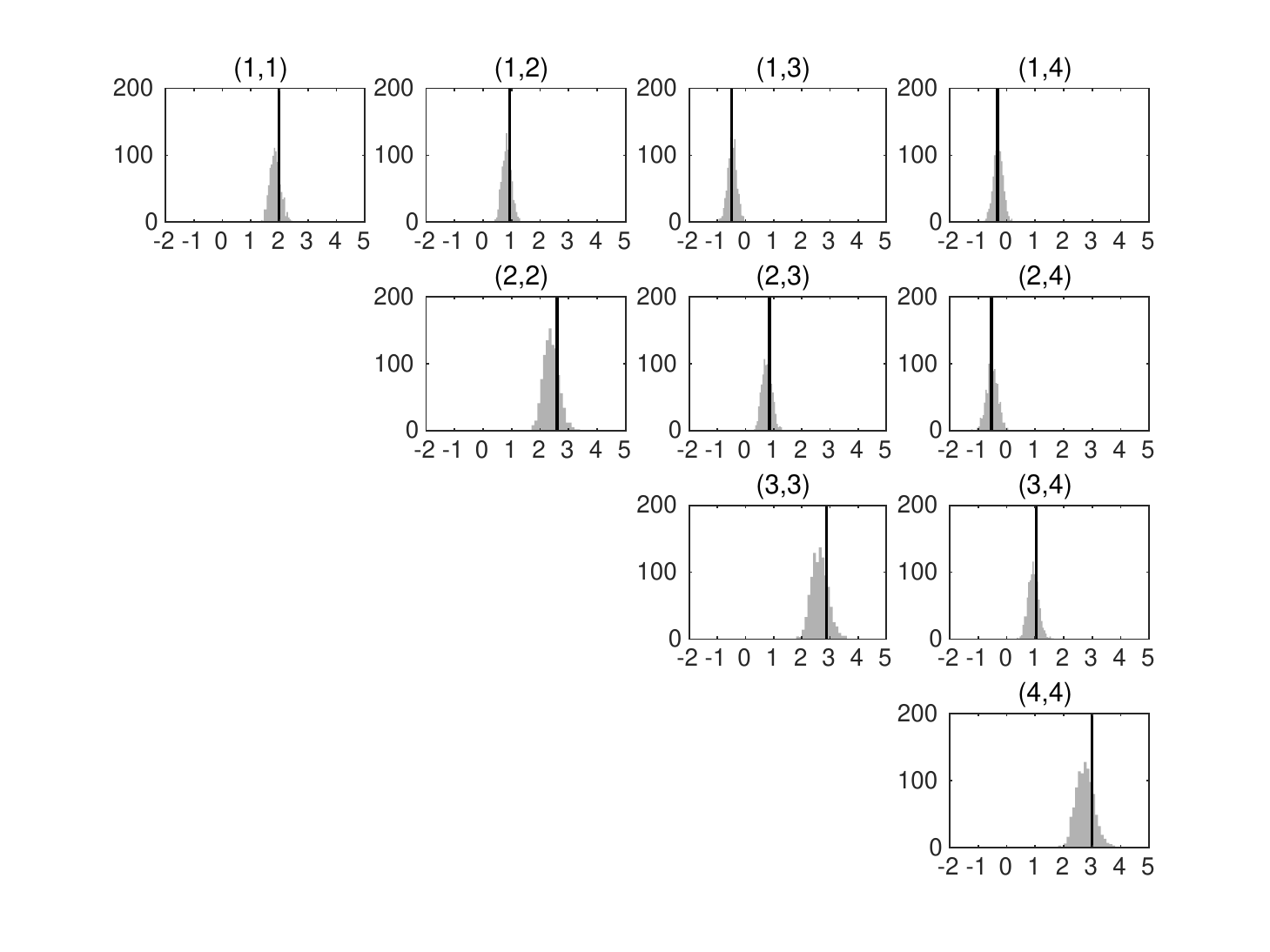}
\par\end{centering}
\centering{}%
\begin{minipage}[t]{0.8\columnwidth}%
Note: The shaded areas indicate the histograms of the OLS estimates
of $\boldsymbol{\Sigma}$. The vertical lines trace the corresponding
true values.%
\end{minipage}
\end{figure}

\clearpage{}

\begin{figure}
\caption{OLS estimates of $\boldsymbol{\Sigma}$ (2)}

\begin{centering}
\includegraphics{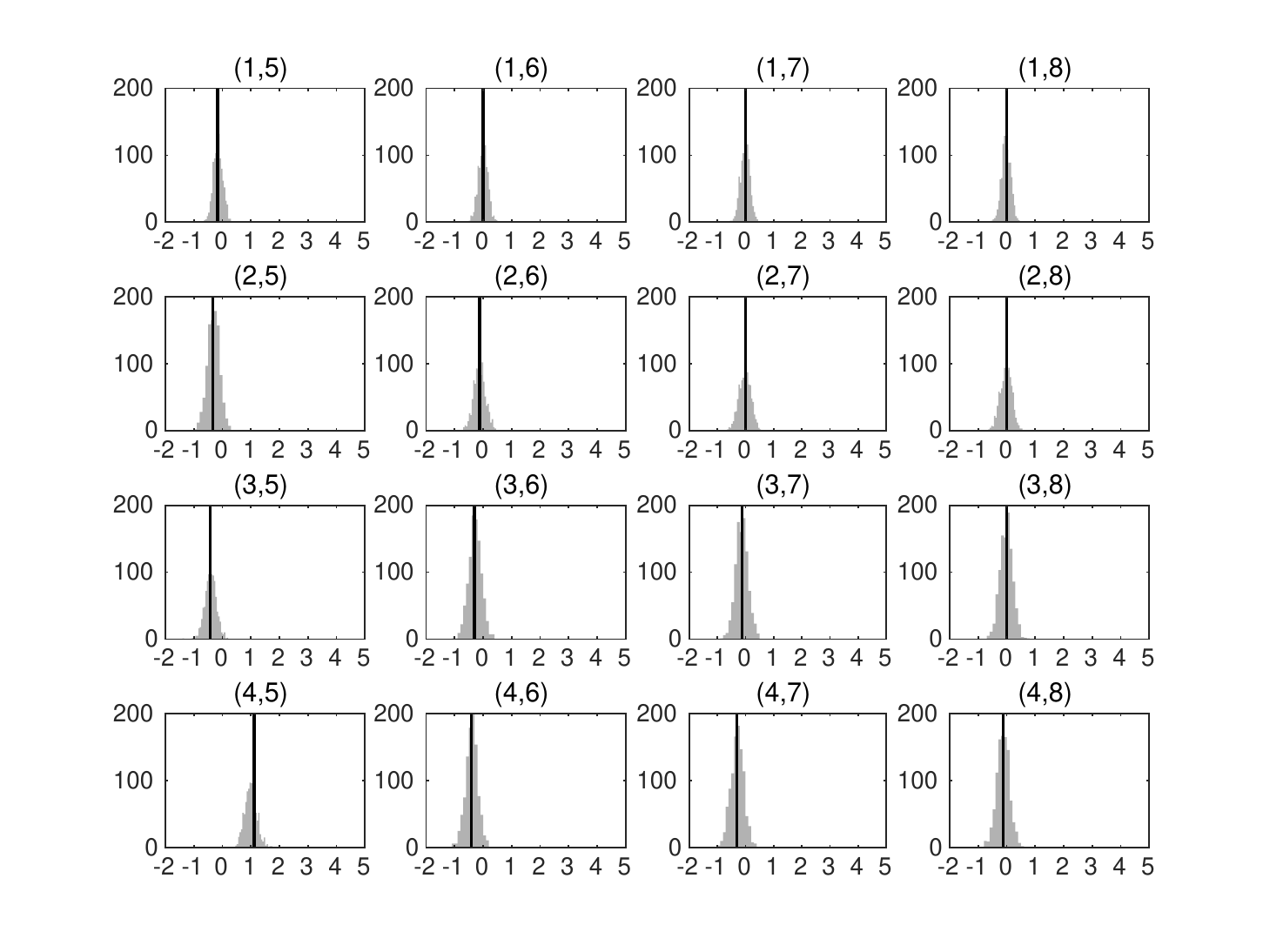}
\par\end{centering}
\centering{}%
\begin{minipage}[t]{0.8\columnwidth}%
Note: The shaded areas indicate the histograms of the OLS estimates
of $\boldsymbol{\Sigma}$. The vertical lines trace the corresponding
true values.%
\end{minipage}
\end{figure}

\clearpage{}

\begin{figure}
\caption{OLS estimates of $\boldsymbol{\Sigma}$ (3)}

\begin{centering}
\includegraphics{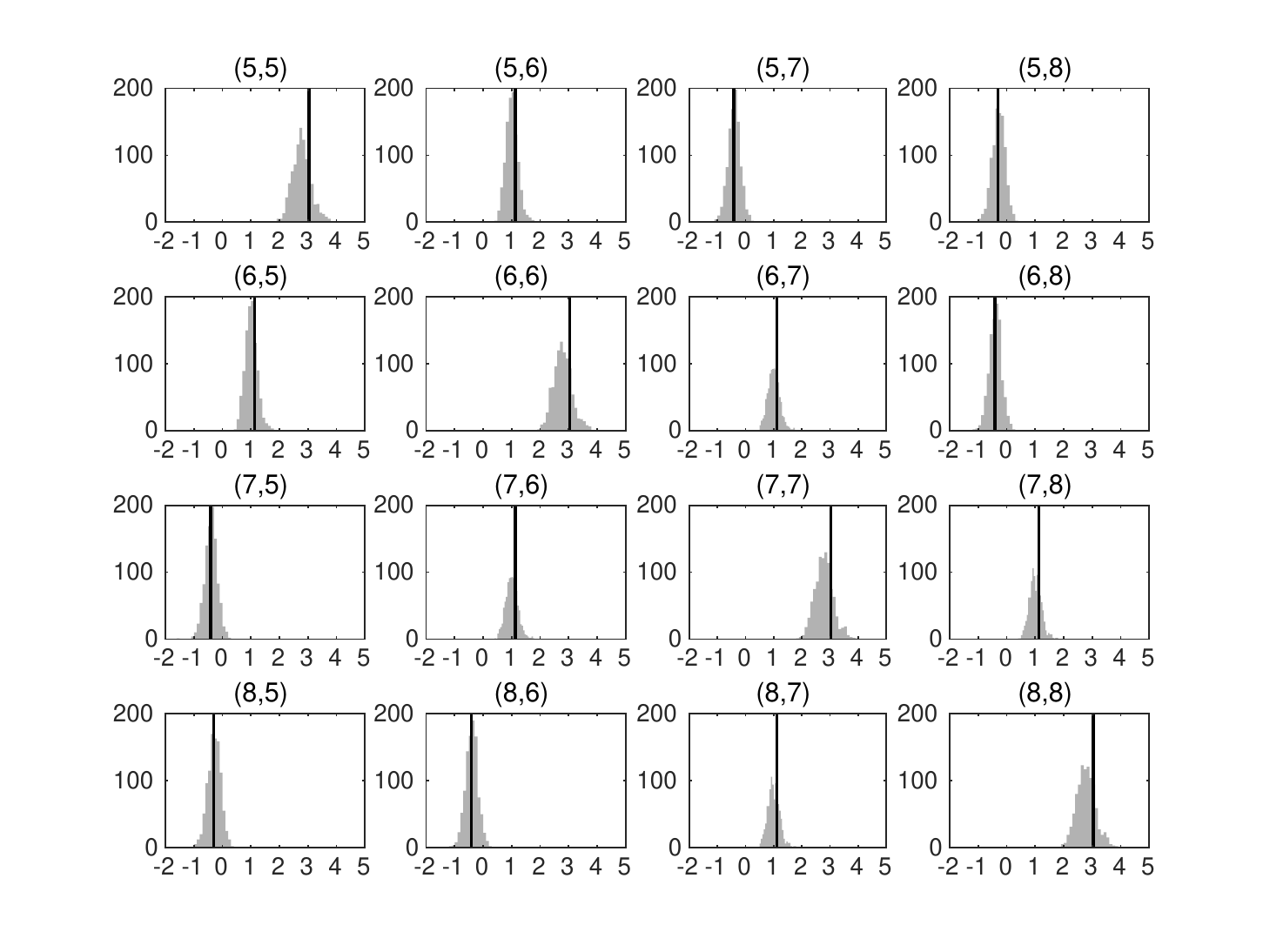}
\par\end{centering}
\centering{}%
\begin{minipage}[t]{0.8\columnwidth}%
Note: The shaded areas indicate the histograms of the OLS estimates
of $\boldsymbol{\Sigma}$. The vertical lines trace the corresponding
true values.%
\end{minipage}
\end{figure}

\clearpage{}

\begin{figure}
\caption{OLS estimates of $\boldsymbol{\gamma}_{IRF}$}

\begin{centering}
\includegraphics{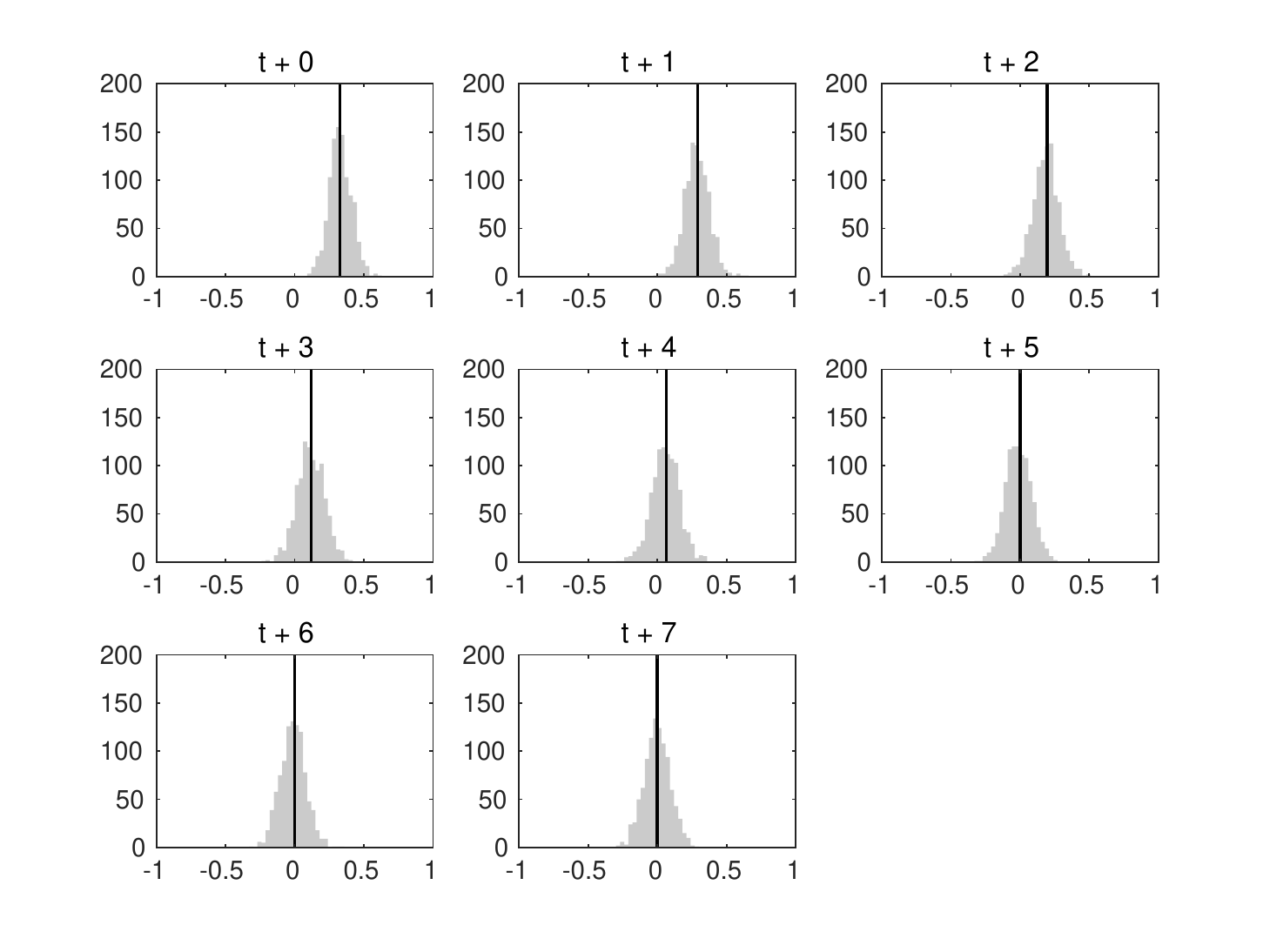}
\par\end{centering}
\centering{}%
\begin{minipage}[t]{0.8\columnwidth}%
Note: The shaded areas indicate the histograms of the OLS estimates
of $\boldsymbol{\gamma}_{IRF}$. The vertical lines trace the corresponding
true values.%
\end{minipage}
\end{figure}

\clearpage{}

\begin{figure}
\caption{Posterior mean estimates of $\boldsymbol{\Sigma}$ (1)}

\begin{centering}
\includegraphics{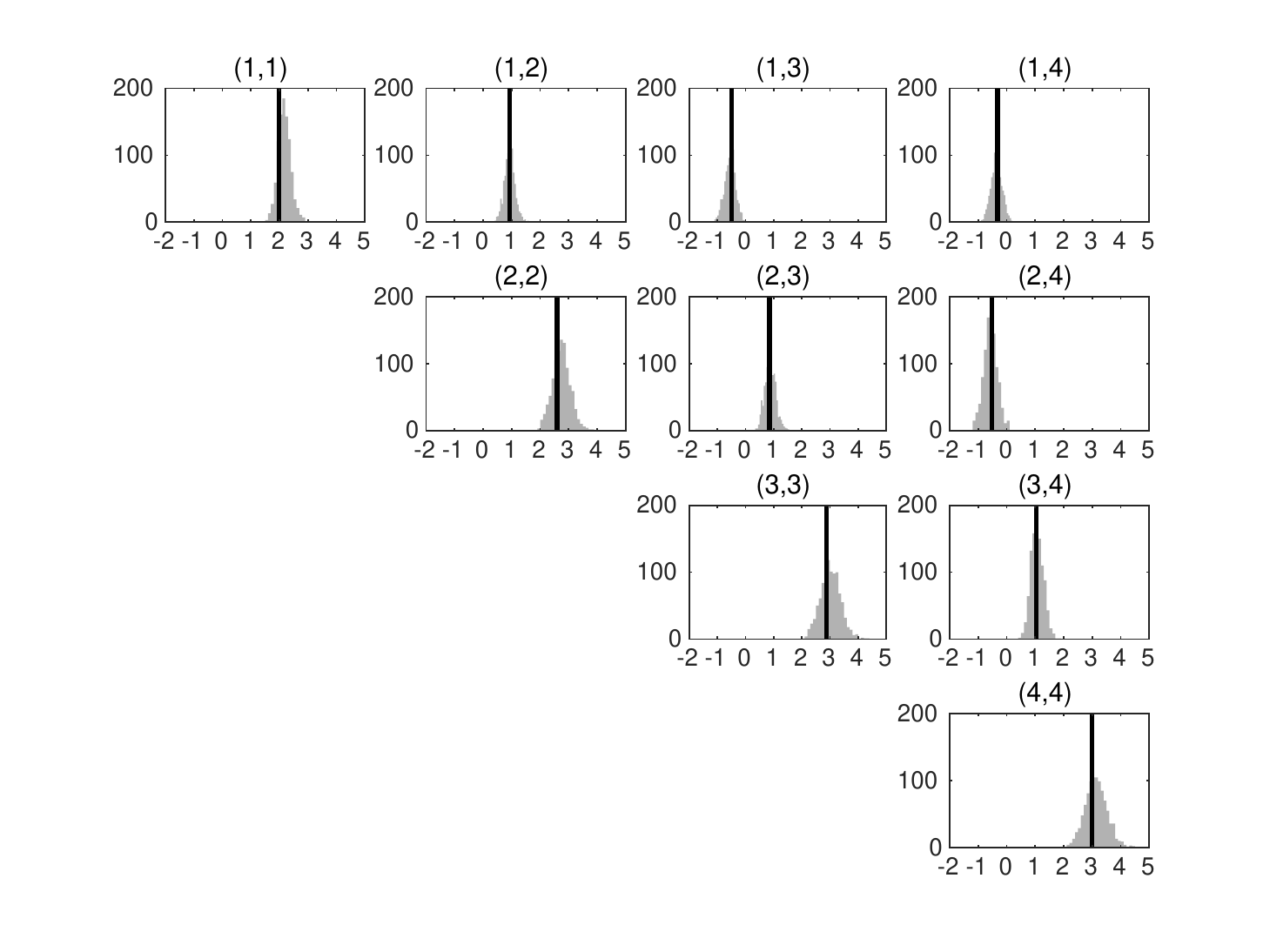}
\par\end{centering}
\centering{}%
\begin{minipage}[t]{0.8\columnwidth}%
Note: The shaded areas indicate the histograms of the posterior mean
estimates of $\boldsymbol{\Sigma}$. The vertical lines trace the
corresponding true values.%
\end{minipage}
\end{figure}
\clearpage{}

\begin{figure}
\caption{Posterior mean estimates of $\boldsymbol{\Sigma}$ (2)}

\begin{centering}
\includegraphics{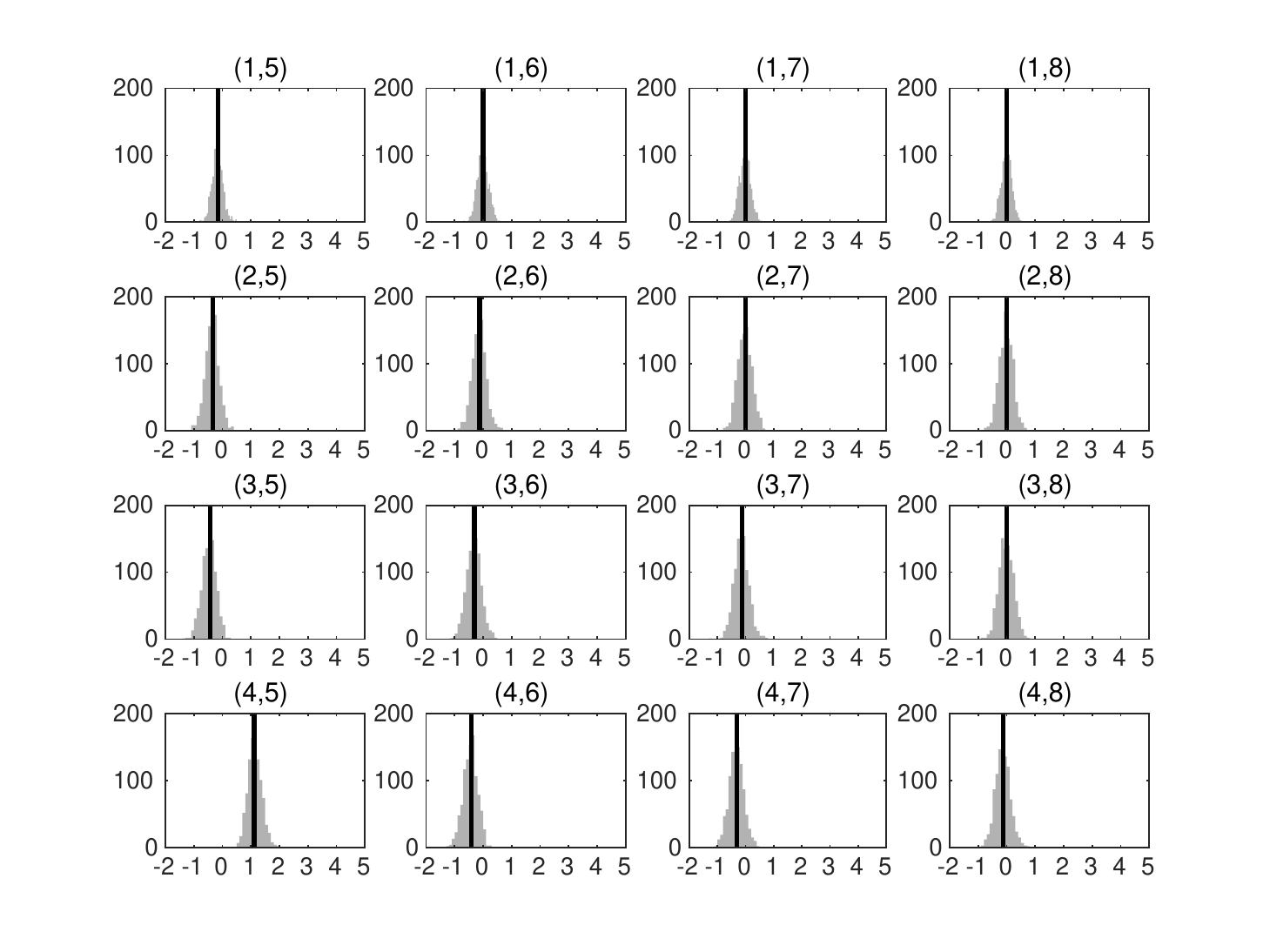}
\par\end{centering}
\centering{}%
\begin{minipage}[t]{0.8\columnwidth}%
Note: The shaded areas indicate the histograms of the posterior mean
estimates of $\boldsymbol{\Sigma}$. The vertical lines trace the
corresponding true values.%
\end{minipage}
\end{figure}

\clearpage{}

\begin{figure}
\caption{Posterior mean estimates of $\boldsymbol{\Sigma}$ (3)}

\begin{centering}
\includegraphics{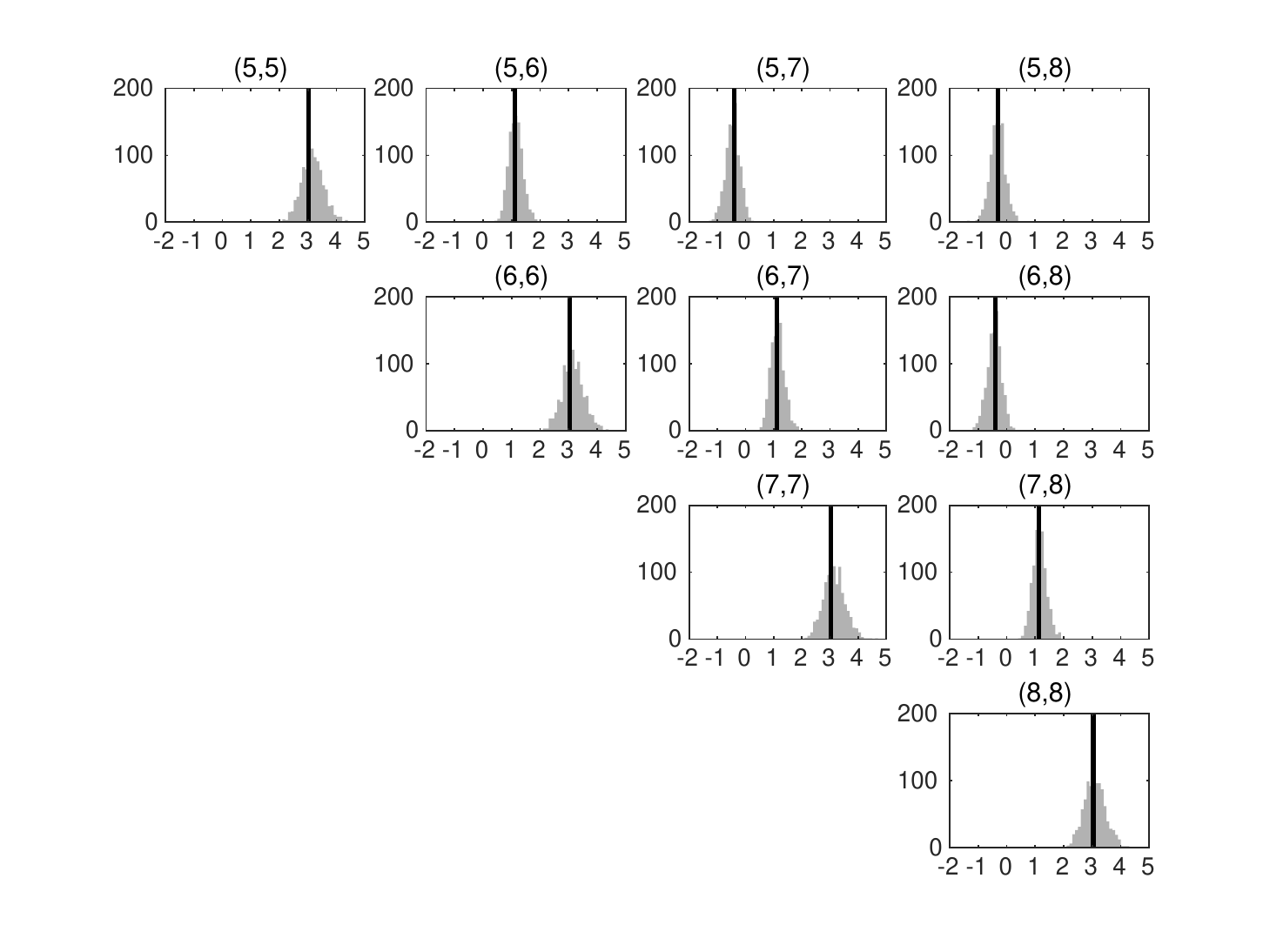}
\par\end{centering}
\centering{}%
\begin{minipage}[t]{0.8\columnwidth}%
Note: The shaded areas indicate the histograms of the posterior mean
estimates of $\boldsymbol{\Sigma}$. The vertical lines trace the
corresponding true values.%
\end{minipage}
\end{figure}

\clearpage{}

\begin{figure}
\caption{Posterior mean estimates of $\boldsymbol{\gamma}_{IRF}$}

\begin{centering}
\includegraphics{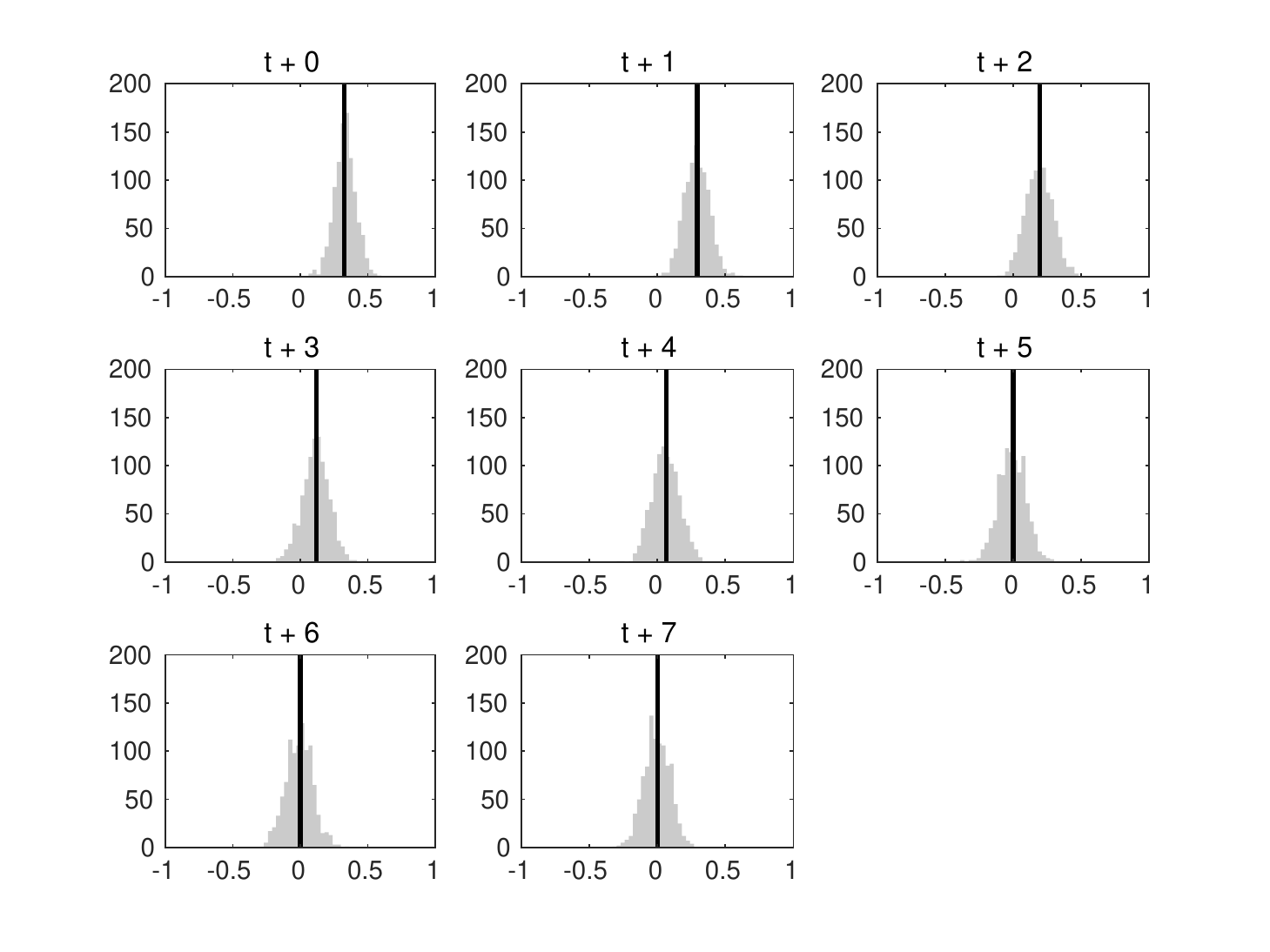}
\par\end{centering}
\centering{}%
\begin{minipage}[t]{0.8\columnwidth}%
Note: The shaded areas indicate the histograms of the posterior mean
estimates of $\boldsymbol{\gamma}_{IRF}$. The vertical lines trace
the corresponding true values.%
\end{minipage}
\end{figure}

\end{document}